\documentstyle[epsf]{mn}

\newif\ifAMStwofonts

\newcommand{\sige}{{\mbox{$\sigma_{\!{\rm\tiny RA}}$}}}
\newcommand{\sigh}{{\mbox{$\sigma_{\!{\rm\tiny Dec}}$}}}
\newcommand{\rmoon}{{\mbox{$r_{\rm\tiny Moon}$}}}
\newcommand{\hr}{$^{\rm h}$}
\newcommand{\mn}{$^{\rm m}$}
\newcommand{\rs}{$^{\rm s}$}
\newcommand{\TT}{$\Delta T/T$}

\def\et{ et~al.~}

\ifoldfss
  \ifCUPmtlplainloaded \else
    \NewTextAlphabet{textbfit} {cmbxti10} {}
    \NewTextAlphabet{textbfss} {cmssbx10} {}
    \NewMathAlphabet{mathbfit} {cmbxti10} {} 
    \NewMathAlphabet{mathbfss} {cmssbx10} {} 
  \fi
  \ifAMStwofonts
    \ifCUPmtlplainloaded \else
      \NewSymbolFont{upmath} {eurm10}
      \NewSymbolFont{AMSa} {msam10}
      \NewMathSymbol{\upi}     {0}{upmath}{19}
      \NewMathSymbol{\umu}     {0}{upmath}{16}
      \NewMathSymbol{\upartial}{0}{upmath}{40}
      \NewMathSymbol{\leqslant}{3}{AMSa}{36}
      \NewMathSymbol{\geqslant}{3}{AMSa}{3E}

       \let\le=\leqslant
       \let\ge=\geqslant
    \fi
  \fi
\fi 

\ifnfssone
  \newmathalphabet{\mathit}
  \addtoversion{normal}{\mathit}{cmr}{m}{it}
  \addtoversion{bold}{\mathit}{cmr}{bx}{it}
  \newmathalphabet{\mathbfit} 
  \addtoversion{normal}{\mathbfit}{cmr}{bx}{it}
  \addtoversion{bold}{\mathbfit}{cmr}{bx}{it}
  \newmathalphabet{\mathbfss} 
  \addtoversion{normal}{\mathbfss}{cmss}{bx}{n}
  \addtoversion{bold}{\mathbfss}{cmss}{bx}{n}
  \ifAMStwofonts
    \ifCUPmtlplainloaded \else
      \UseAMStwoboldmath
      \makeatletter
      \new@mathgroup\upmath@group
      \define@mathgroup\mv@normal\upmath@group{eur}{m}{n}
      \define@mathgroup\mv@bold\upmath@group{eur}{b}{n}
      \edef\UPM{\hexnumber\upmath@group}
      \new@mathgroup\amsa@group
      \define@mathgroup\mv@normal\amsa@group{msa}{m}{n}
      \define@mathgroup\mv@bold\amsa@group{msa}{m}{n}
      \edef\AMSa{\hexnumber\amsa@group}
      \makeatother
      \mathchardef\upi="0\UPM19
      \mathchardef\umu="0\UPM16
      \mathchardef\upartial="0\UPM40
      \mathchardef\leqslant="3\AMSa36
      \mathchardef\geqslant="3\AMSa3E

       \let\le=\leqslant
       \let\ge=\geqslant
    \fi
  \fi
\fi 

\ifnfsstwo
  \DeclareMathAlphabet{\mathbfit}{OT1}{cmr}{bx}{it}
  \SetMathAlphabet\mathbfit{bold}{OT1}{cmr}{bx}{it}
  \DeclareMathAlphabet{\mathbfss}{OT1}{cmss}{bx}{n}
  \SetMathAlphabet\mathbfss{bold}{OT1}{cmss}{bx}{n}
  \ifAMStwofonts
    \ifCUPmtlplainloaded \else
      \DeclareSymbolFont{UPM}{U}{eur}{m}{n}
      \SetSymbolFont{UPM}{bold}{U}{eur}{b}{n}
      \DeclareSymbolFont{AMSa}{U}{msa}{m}{n}
      \DeclareMathSymbol{\upi}{0}{UPM}{"19}
      \DeclareMathSymbol{\umu}{0}{UPM}{"16}
      \DeclareMathSymbol{\upartial}{0}{UPM}{"40}
      \DeclareMathSymbol{\leqslant}{3}{AMSa}{"36}
      \DeclareMathSymbol{\geqslant}{3}{AMSa}{"3E}

       \let\le=\leqslant
       \let\ge=\geqslant
    \fi
  \fi
\fi 

\ifCUPmtlplainloaded \else
  \ifAMStwofonts \else 
    \def\upi{\pi}
    \def\umu{\mu}
    \def\upartial{\partial}
  \fi
\fi

\label{firstpage}
\title{CMB observations with the Jodrell Bank -- IAC interferometer at 33\,GHz}

\author[]{S.R.Dicker$^{1,2}$, S.J. Melhuish$^1$, R.D. Davies$^1$, C.M.
Gutierrez$^3$,
R. Rebolo$^3$, \newauthor D.L.Harrison$^1$, R.J. Davis$^1$, A.Wilkinson$^1$, R.J.
Hoyland$^3$, R.A. Watson$^1$\\
$^1$University of Manchester, Nuffield Radio Astronomy Laboratories,
Jodrell Bank, Macclesfield, Cheshire  SK11 9DL UK\\
$^2$David Rittenhouse Labs., Department of Physics and Astronomy, 209 S. 33$^{rd}$
Street, University of Pennsylvania, Philadelphia,\\ 
PA, USA\\
$^3$Instituto de Astrofisica de Canarias, 38200 La Laguna, Tenerife, Canary
Islands, Spain}

\pagerange{\pageref{firstpage}--\pageref{lastpage}}
\pubyear{2000}

\begin{document}

\maketitle


\begin{abstract}

The paper presents the first results obtained with the Jodrell~Bank~--~IAC two-element
33\,GHz interferometer. The instrument was designed to measure the level of the
Cosmic Microwave Background (CMB) fluctuations at angular scales of $
1\degr-2$\degr. The observations analyzed here were taken in a strip of the sky at
Dec=+41\degr~with an element separation of 16.7$\,\lambda$, which gives a
maximum sensitivity to $\sim 1\fdg6$ ~structures on the sky. The data
processing and calibration of the instrument are described.  The
sensitivity achieved in each of the two channels is 7 $\mu$K per resolution
element. A reconstruction of the sky at Dec=+41\degr ~ using a maximum
entropy method shows the presence of structure at a high level of significance.
A likelihood analysis, assuming a flat CMB spatial power spectrum,
gives a best estimate of the level of CMB fluctuations of
$\Delta T_\ell=43^{+13}_{-12}$ $\mu$K for the range $\ell=109\pm19$;
the main uncertainty in this result arises from sample variance. 
We consider that the contamination from the Galaxy is small.
These results represent a new determination of the CMB power spectrum on
angular scales where previous results show a large scatter;
our new results are in agreement with the theoretical predictions
of the standard inflationary cold dark matter models.

\end{abstract}


\section{Introduction}

Since the COBE DMR  detection of anisotropy (Smoot \et 1992)
and the direct observation of individual structures (for example Hancock \et 1994) on the
Cosmic Microwave Background (CMB), many other
detections and upper limits have been reported on angular scales ranging from
15\degr\ to a few arc-minutes (see Lineweaver (1998), Tegmark (1998) for recent
reviews). Despite all of this effort, the shape of the CMB power spectrum is still
poorly defined. There is some observational evidence that the power
spectrum has a peak centred around spherical harmonics $\ell$=200 (Hancock \et 1998)
supporting the cold dark
matter models which predict this peak as a consequence of the acoustic
oscillations in the primordial plasma. Its position, amplitude and
height depend on fundamental cosmological
parameters such as the density of the universe $\Omega$, the density of the
baryonic component $\Omega _b$ and the Hubble constant $H_o$; this explains
the large observational effort dedicated to the determination
of the height and location in $\ell$ space of this peak.

In this paper we analyze the first results of the Jodrell~Bank~--~IAC 33\,GHz
interferometer experiment. The aim of the project is to measure the level of CMB
fluctuations in the range $\ell =100-200$. The data presented here were taken
at the Teide Observatory, Tenerife, between 4 April 1997 and 9 March
1998 using the low spacing configuration which corresponds to an
angular spherical harmonic $\ell \sim 100$. The paper is organized as follows:
Sections~\ref{int} and \ref{basic} present a brief description of the
instrument and the data processing respectively. The methods used for
calibration are explained in Section~\ref{cal}. The reconstruction of
the sky signal using a Maximum Entropy Method is presented in
Section~\ref{MEM}. Sections~\ref{like} and~\ref{dis} analyze
statistically the data using a likelihood analysis and discuss
the possible contribution of foregrounds. The conclusions and future 
programme are presented in Section ~\ref{con}.

\section{The 33\,GHz Interferometer}
\label{int}
The full description of the instrument configuration and
observing strategy can be found in Melhuish \et (1998). A brief summary
of the main parameters of the instrument will now be given.
The interferometer consists of two  horn-reflector antennas  positioned to
form a single E--W baseline of length 152\,mm for the observations presented
here. Observations are made at constant
declination using the rotation of the Earth to ``scan''
24\hr\ in RA each day. The horn polarization is horizontal -- parallel with the scan direction.
The observations analyzed here were made at
Dec=+41\degr; further observations at other declinations are now in
progress. There are two data outputs representing the cosine and the
sine parts of the complex interferometer visibility. The
operating frequency range is 31--34\,GHz,  near a local minimum in the
atmospheric emission. The low level of precipitable water vapour, which is
typically around 3\,mm at Teide Observatory, Iza\~{n}a, permits the
collection of high quality data limited by the receiver noise for over
80 per cent of the time.  In these good weather conditions the system has an
RMS noise of 220\,$\mu$K in a 2-minute integration. 
The receivers employ cryogenically cooled, low noise, HEMT
amplifiers, and have a bandwidth of $\sim$3\,GHz. To achieve the
sensitivity required to measure CMB anisotropy  ($\sim10\,\mu{\rm K}$ per
resolution element), repeated observations are stacked together as
explained in Section~\ref{basic}.
The measured beam
shape of the interferometer is well approximated by a Gaussian with
sigmas of \sige=2\fdg25$\pm$0\fdg03 (in RA) and
\sigh=1\fdg00$\pm$0\fdg02 (in Dec),  modulated by fringes with a period
of $f$=3\fdg48$\pm$0\fdg04 (in RA). This defines the range of sensitivity
to the different multipoles $\ell$ of the CMB power spectrum ($C_{\ell}$) in the range 
corresponding to a maximum sensitivity at $\ell$=109 (1\fdg6) and half
sensitivity at $\Delta \ell=\pm$19. The results of the beam-switching
Tenerife experiments at 5\degr~angular scales (Hancock \et 1997, Guti\'errez \et 1998)
and realistic models for the power spectra of the diffuse Galactic
emission (Lasenby 1996, Davies and Wilkinson 1998) indicate that, for our
frequency and angular scale, Galactic contamination is more than a factor of
10 below the intrinsic CMB fluctuations in a section of the sky at Dec$\sim
41$\degr~at high Galactic latitude (see also Section~\ref{Gal}).

\begin{figure}
\begin{center}
\leavevmode\epsfbox{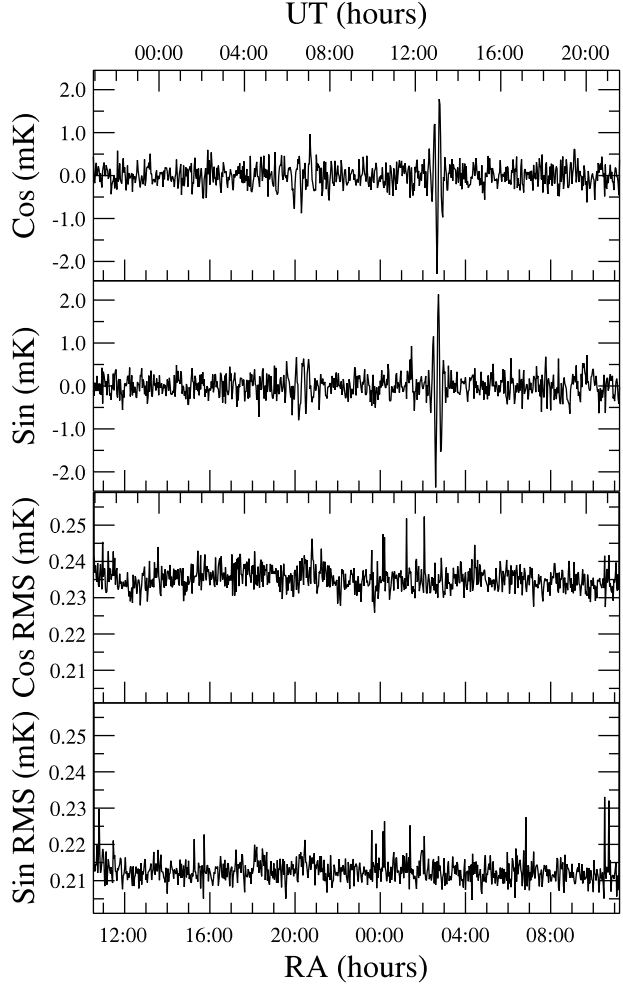}
\caption{
A single interferometer data file at +41\degr declination, covering just over
24 hours. Basic processing steps, such as calibration and correction for the
error in correlator phase quadrature, have been performed. Fringes are just
visible at the Galactic plane crossing around 20 hours RA. The feature near
13:00 U.T. is accounted-for by off-axis pick-up of emission from the Sun.
}
\label{dayofdata}
\end{center}
\end{figure}

A known signal (CAL) is periodically injected
into the waveguide connecting the horns to the HEMTs
 allowing a continuous
calibration and concomitant corrections for drifts in the system gain and phase offset.
The data acquisition cycle lasts 30 seconds during which time two 14-second
integrations with CAL off and two 1-second integrations with CAL on are carried out.
 These two integrations are combined to form a 28-second CAL-off integration
(which measures the astronomy) and a 2-second CAL-on integration for the cosine and sine data
channels.  Details of how these data are processed can be found in
Section~\ref{basic}. Each integration is made by averaging 15-ms sub-integrations. 
The scatter in the sub-integrations is entirely due to atmospheric and
system noise since on these time
scales the changes in the astronomical signal are negligible.
 This scatter is used to estimate the RMS
values for the CAL-off and CAL-on integrations.
 Fig.~\ref{dayofdata} shows an example of the CAL-off
and RMS 30-second data taken on 2 May 1997; baselines have been removed,
but no other filtering was applied.

\section{Basic Data processing}
\label{basic}
\subsection{Calibrating the raw data}

As a first step in the analysis, baselines are subtracted
from the CAL-off cosine and sine data and any departure from the quadrature between
the sine and cosine data of both the CAL-off and CAL-on records is corrected. On the
time-scale it takes to carry out a single 30-second
integration cycle, the amplitude and phase response of the instrument
can be considered unchanged. By carrying out a complex division of the
CAL-off data by the CAL-on data all variations in instrumental performance
cancel and calibrated data in units of the amplitude of the CAL signal
are obtained. These data are then converted to physical units  by
multiplying by the measured amplitude of the CAL signal (Section~\ref{cal}).
The procedures for removing the baselines and correcting the quadrature will
now be outlined.

\subsubsection{Baseline removal}

The raw CAL-off data contain a small offset of amplitude 
$\sim$ 1\,mK. Daily variations in this offset will appear as extra noise in
the stacked data. A high-pass Gaussian filter (i.e. the result of
subtracting a low-pass Gaussian filter applied to the data) was used to
remove the offset. If the width of this filter is too narrow, higher frequency
components of the baseline will remain, thus increasing the noise. Conversely
a filter that is too wide will filter out some of the fringes, reducing the
astronomical signal. The optimum filter is the one that removes as much of the
baseline as possible while leaving the astronomy intact, thus giving the
largest signal-to-noise ratio. Observations of the Moon, used for calibration, showed
that for typical values of the offset the optimum width of the Gaussian filter
was $\sigma$=3\fdg2; the reduction in the astronomical signal is negligible
and has no effect on the temperature scale as survey and calibration data are
scaled by the same factor.

\subsubsection{Correcting the quadrature}

The output of both channels can be modelled as:
\begin{eqnarray}
\mbox{cosine channel:}  &  A\cos \theta \nonumber \\
\mbox{sine channel:}    &  A(1+\delta)\sin(\theta + \Delta) \nonumber
\end{eqnarray}
where $\delta$ is the relative difference in sensitivity between
the two channels and $\Delta$ is the departure from quadrature.
Provided $\delta$ and $\Delta$ are both small, the data can be brought
back into quadrature, with negligible loss of sensitivity  using:
\begin{equation}
\label{matrix}
\left[\begin{array}{c}\mbox{cosine}\\\\\mbox{sine} \end{array}\right]
=
\left[\begin{array}{cc} 1 & 0 \\\\
 -\tan\Delta   &   {\frac{1}{(1+\delta)\cos\Delta}}
 \end{array}\right] 
\left[\begin{array}{c}\mbox{cosine $^\prime$}\\\\\mbox{sine $^\prime$}
\end{array}\right] 
\end{equation}
Moon observations showed 
$\delta$ to be negligible; while $\Delta$ was found to be 
6\fdg1, small enough to be corrected using equation~\ref{matrix}.

\subsection{Editing and quality control of the data}

After this relative calibration, the data were re-binned into 2 minute (RA) 
bins
and edited to remove periods of bad
weather, spikes and times when the Sun may have contaminated the data.
All data points outside a $\pm$650\,$\mu$K range ($\sim$3 times the
instrumental RMS value) were removed. A visual inspection was carried
out and any regions of data with more than $\sim$30 per cent of the
points deleted were discarded. 
 
Every 10 days data were stacked, combining all points with the same
UT; with this stacking  only features due to the Sun will remain in the
data. The positions of these features were noted and these areas were
deleted in the original data files used to make that stack.  Any solar
features below the noise in a 10-day UT stack can be neglected as 
any signals not fixed in RA will be smeared out in the final RA stacks.
For most of the year only the midday Sun transit 
was visible, so only data within $\pm$1 hour of this feature were
removed. The exception to this was around mid-summer when all data
within $\pm$8 hours of the Sun transient were deleted.

In a single day, with the exceptions of the midday Sun crossing and the
Cygnus region in the Galactic plane, the noise is much greater than any
astronomical signal. Consequently each data point should be drawn from a normal
distribution with a sigma equal to the RMS record which contains
information on time-scales less than 30 seconds. Spikes, weather and
variations in the offset which occur on large time-scales result in data
points with a larger scatter than their RMS records
suggest. 

\begin{figure}
\leavevmode\epsfbox{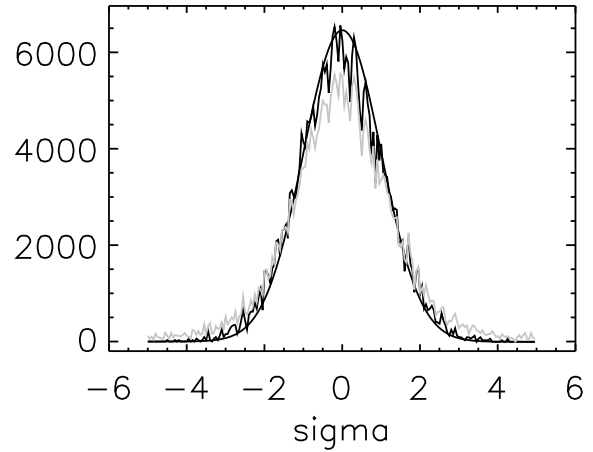}
   \caption{\label{rms} The statistics of the data.  This plot is a histogram
   of 12 days of data divided by their RMS records.  The gray and black 
   lines are the unedited and edited data respectively, while the smooth line
   is the expected $\sigma = 1$ Gaussian distribution.}
\end{figure}

Fig.~\ref{rms} shows a histogram of the distribution of data points,
normalized by dividing through by the mean RMS level. The darker
and lighter curves correspond to the edited and unedited data respectively
(in both cases the Cygnus region RA=20\hr30\mn$\pm$ 1\hr and the Sun transit
have been removed from the data). The smooth line shows the expected
distribution if the noise in the data was accurately reflected in the
RMS record. This measures noise on timescales $< 30\,$s, and is
dominated by receiver noise. Both curves follow normal distributions showing that all
experimental errors are Gaussian; however the unedited data have a
broader distribution showing the effects of weather and
other sources of additional noise. The distribution of the edited data is
close to the expected curve demonstrating that the data have been edited
properly and are limited by system noise alone. 

\subsection{The stability of CAL}

\begin{figure*}
\leavevmode\epsfbox{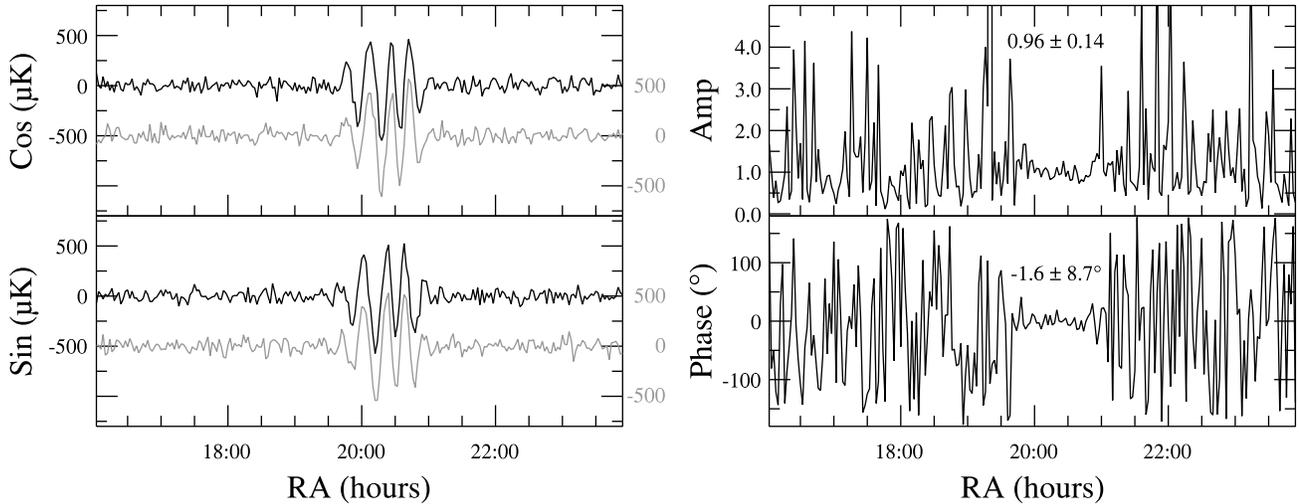}
\caption{\label{stab}The stability of CAL. The two stacks shown on the left
(April/May 1997 in black and August 1997 grey)
correspond to approximately 20 days each. On the right is shown
the complex division of these two stacks over the
Cygnus region (RA=20\hr00 to 20\hr45), where the signal is much larger
than the noise. The amplitude is $\sim$1 and the phase
$\sim$0\degr\ showing that there has been no change in the 
Cygnus region, and hence in the amplitude and phase of CAL (see main text).  
}
\end{figure*}

The data from individual scans, following the processing steps outlined
above, will be collected together in a 24-hour {\em stack}. The value at
each RA position in the stack is given by the mean value at that RA
from the contributing scans. This procedure relies on the CAL source being
stable.
Variations in the amplitude of CAL would increase the noise in the data,
and changes in phase  by more than a few degrees, as well as averaging out
the noise, would smear out the astronomical signal when the data are
stacked. To check the stability of CAL, two independent stacks were made,
one using data from April/May 1997 and the other using data from August
1997. In areas where the astronomical signal dominates over the noise, the
result of the complex division\footnote{The cosine and sine channels are
the real and imaginary parts of a complex visibility. We may equivalently
express data in terms of their amplitude (which is always positive) and phase.}
of one of these stacks by the other should
have an amplitude of 1 and a phase of 0\degr~if CAL is perfectly stable.
Using the Cygnus region shown in Fig.~\ref{stab} as an astronomical reference
it was demonstrated
that on a time-scale of 14 weeks CAL is stable to approximately 4 per cent
in amplitude and 2\degr\ in phase. This process was repeated using pairs of 
stacks made from data taken at
epochs separated by 5 weeks and 6 months. On the 5-week time-scale CAL
was stable to better than 3 per cent in amplitude and 2\degr\ in
phase, while on the 6 month time scale the values were 5 per cent and
4\degr\ respectively. These results
show that CAL is highly stable and that data taken many months apart
can be stacked with negligible error. 

\subsection{The Dec=+41\degr\ stack.}

\begin{center}
\begin{figure*}
\leavevmode\epsfbox{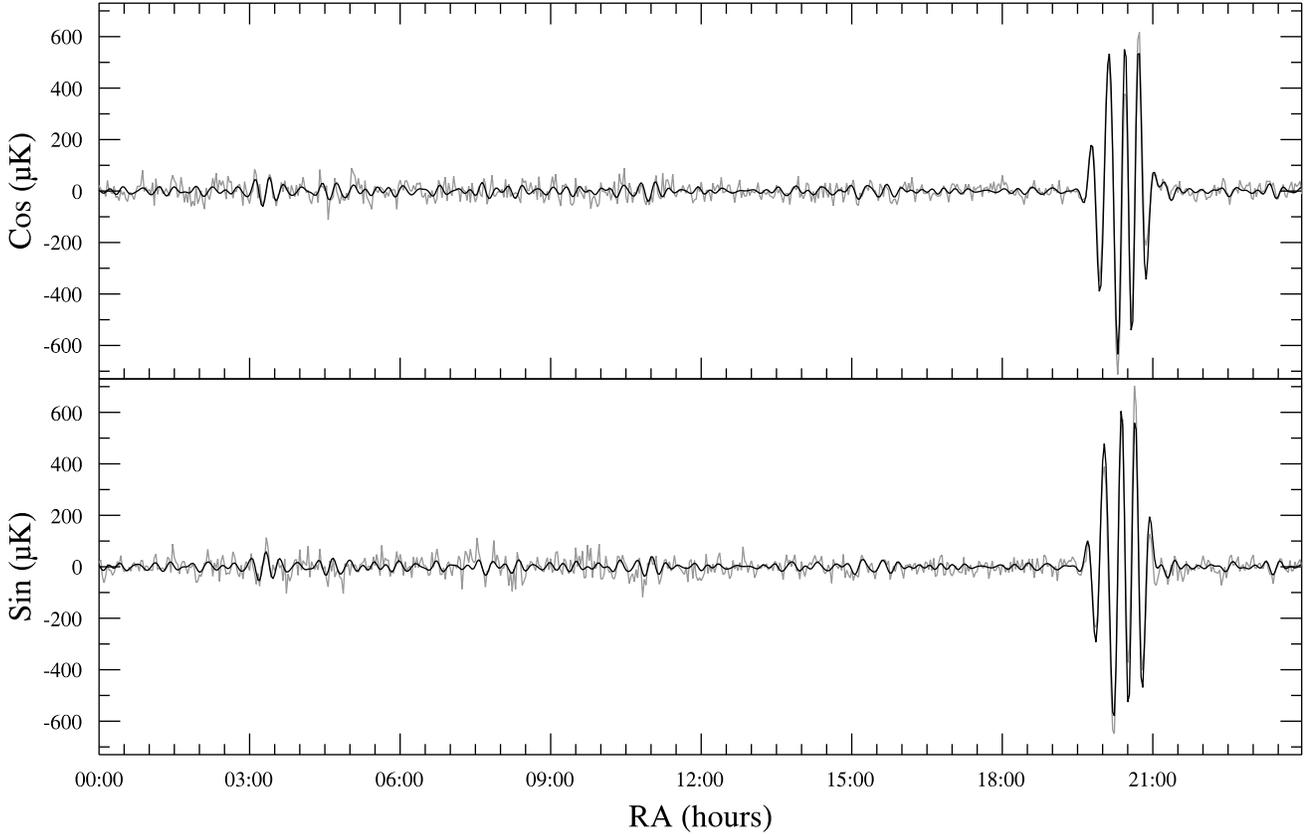}
\caption{\label{Fig:MEMstack}
The stack of the data collected at Dec=+41\degr. 
The diagram shows the cosine and sine visibility data.
The MEM-reduced data stack (heavy line) is
plotted over the basic 2-minute data stack (light line).}
\end{figure*}
\end{center}

A total of 100 days of usable data were collected at Dec=+41\degr.
These were combined into the stack shown in Fig.~\ref{Fig:MEMstack}.
The number of observations at each RA varies between 127 and
60 due to the editing out of the Sun. The RMS at each RA is that
expected from the RMS on a good day divided by the square root of the
number of days of data at that RA, further demonstrating the
effectiveness of our editing. The RMS values of the cosine data are marginally
higher than the sine data. We attribute this to the asymmetric nature
of the sine fringes being slightly better at cancelling atmospheric noise
than the symmetric cosine interferometer pattern.

One feature that stands out is the Cygnus region between RA=19\hr30\mn and
21\hr30\mn.  When plotted in amplitude and phase, two peaks at
RA=20\hr18\mn\ and 20\hr40\mn\ are clearly distinguishable. Similar peaks
can be seen in our 5\,GHz survey of the same region (Asareh 1997).
Most of this observed structure is due to diffuse Galactic
free-free emission; however
the radio galaxy Cygnus~A (RA=19\hr57\mn\ Dec=40\degr\ 36$'$), which has an
expected amplitude at Dec=41\degr of 200$\mu$K, can just be distinguished as the
slight bulge on the low RA side of the Cygnus region.

\section{Temperature calibration}
\label{cal}

The amplitude of the CAL signal was measured by the observation of
known astronomical sources and also by using hot and cold loads. 
Astronomical calibration is more robust than
calibration  based on the response to hot/cold loads as it
automatically takes into account systematic errors such as the
efficiency of the horn feeds and any attenuation in the atmosphere.
Furthermore with practical hot loads allowances must be made for
saturation of the receivers. Conversely, astronomical calibration is
limited by the shortage of accurate data for suitable sources at high
frequencies.  However, an error in the assumed flux of a calibration
source can be easily corrected, whereas the systematic errors occurring
with hot/cold load calibration cannot.  As a result CAL was measured using
astronomical calibration and calibration with hot/cold loads was used
as an additional check.


The large ($\sim 2$\degr $\times 5$\degr) size of the primary beam of the
interferometer results in low sensitivity to point sources, so only the brightest point
sources can be used as astronomical calibrators. At 33\,GHz the three
brightest sources in the sky are the Sun, the Moon and Tau~A (the Crab
nebula), all of which are small in comparison to our  beam size. Of
these, the Sun is too bright, causing saturation of the receivers,
making it unsuitable as a calibrator. By contrast, observations of
Tau~A showed that it had a peak amplitude only $5-6$ times the noise in
a 2-minute integration meaning that many days of observations would
have to be used in order to obtain an accurate calibration. As is shown
below, the power received from the Moon corresponds to an
antenna temperature  in the range 2--4\,K, which is less than half the
power received from the sky ( $\sim$ 10\,K due to the CMB and the atmosphere) 
and so saturation of the
receivers is not a significant factor. However a 2--4\,K signal is large
enough to give signal-to-noise ratios of $\sim$6000 in a single observation.
Consequently the Moon  was used as our primary astronomical calibrator
and observations of Tau~A were used to confirm the calibration
obtained.

\subsection{Calibration using the Moon}

The Moon was modelled as a uniform disk of radius $\rmoon$\ and a 33\,GHz
brightness temperature, $T_{\rm b}$ given by:

\begin{equation}
T_{\rm b}=214 + 36 \cos (\phi - \epsilon) \:\:\mbox{K}
\end{equation}
where $\phi$\ is the phase of the Moon (measured from full Moon) and
$\epsilon$ = 41\degr\ is a phase offset caused by the finite thermal
conductivity of the Moon (Hagfors 1970). The expected antenna
temperature, $T_{\rm E}$, can then be found by integrating over the disk of
the Moon, multiplied by the normalized interferometer beam function:
\begin{eqnarray} \label{dilute} 
T_{\rm E} & = & \frac{T_{\rm b}}{2\pi\sige\sigh} \\ \nonumber
& \times & \int_0^{2\pi}\!\!\!\!\!\int_0^\rmoon \!\!\!\!\!\!\!\!\!\! \exp\left(
{-\frac{x^2}{2\sige^2}} {-\frac{y^2}{2\sigh^2}}\right)\; \cos \left(\frac{2\pi
x}{f}\right) \; {\rm d} \Omega \end{eqnarray}
where $x = \theta\cos\phi$\ and $y = \theta\sin\phi$.  \sige\ and
\sigh\ are the RA and Dec beam sigmas  and $f$ is the fringe spacing
(see Section~\ref{int}).

\begin{figure}
\leavevmode\epsfbox{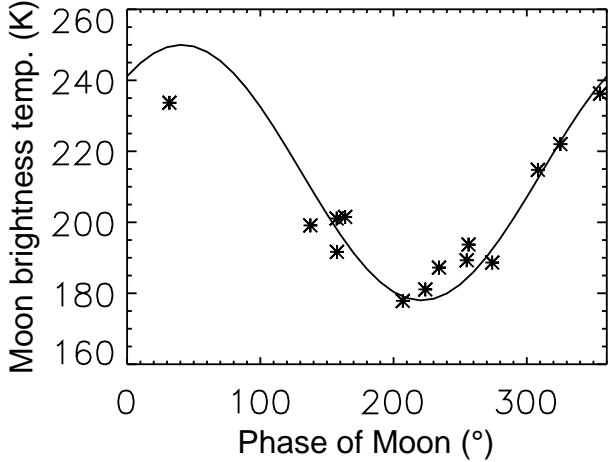}
\caption{\label{moonphase}
Measured value of the brightness
temperature of the Moon.  Each of the
observations has been calibrated using CAL=15.2\,K. The solid line is
the prediction of the model based on data given by Hagfors (1970):
$T_b = 214 + 36 \cos (\phi - 41) $ in Kelvins. Phase is measured
from full Moon.
}
\end{figure}

Regular observations of the Moon were made and the data were processed
as described in Section~\ref{basic}. For each observation,
equation~\ref{dilute} was evaluated numerically and an amplitude for
CAL found so that the amplitude of the Moon in the processed data was
the same as the predicted value. Using 14 observations of the Moon, an
average amplitude for CAL of 15.2$\pm$1.0\,K was found.  The error is
made up from a 3.4 per cent error in the measurements and an estimated
6 per cent in the data presented by Hagfors. 
A small measurement error in beam area, such as $\sim 4$ per cent, 
would not affect the overall calibration, since
astronomical signals would be re-scaled by the same factor as the Moon
calibration.
Fig.~\ref{moonphase} shows our observations and how the measured brightness
temperatures of the Moon changed with phase.

\subsection{Calibration using Tau~A} 
\label{tau2}

Tau~A (the Crab Nebula) is a supernova remnant $6' \times 4'$
in diameter. Using the model of Baars \et (1977) and
taking into account a secular decrease of 0.166 per cent per year
(Aller \& Reynolds 1985), at the epoch of observation (1997) Tau~A has a
32.5\,GHz flux density of 356$\pm$40\,Jy. The measured primary beam-shape of
the interferometer was used to calculate an effective area for each antenna
of the
interferometer, and this was used to convert the flux density of Tau~A into an
antenna temperature. For an unpolarized source it was calculated that
the interferometer has a sensitivity of $7.07\,{\rm~\mu K Jy}^{-1}$, so 
taking into account a 6.6 \% polarization of Tau~A at position angle 150\degr
(Mayer \& Hollinger 1968) the interferometer $E$ vector at PA=90\degr\ should
observe Tau~A to have an amplitude of 2.51$\pm$0.28\,mK.

\begin{figure}  
\leavevmode\epsfbox{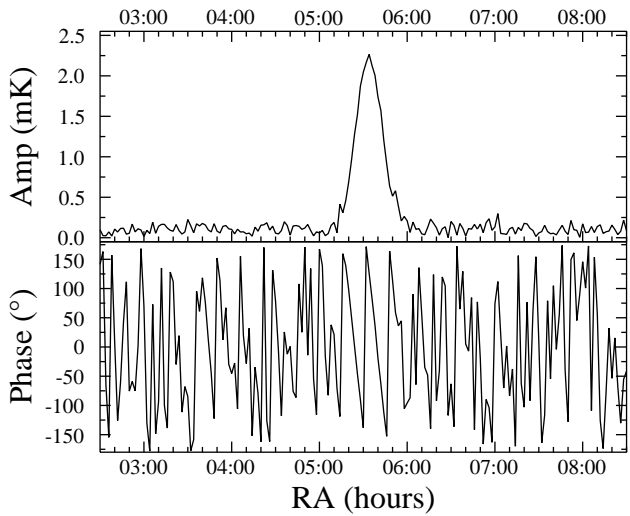}
\caption{\label{Tau}
A stack of 12 days of data at Dec=+22\fdg0, showing Tau A in
amplitude and phase. The amplitude of the peak is 2.51$\pm$0.09\,mK
assuming a flux density of 356\,Jy.
}
\end{figure}

Twelve days of observations at Dec=+22.0\degr\ were processed and combined to
produce the stack shown in Fig.~\ref{Tau}. The measured amplitude of
Tau~A was found by fitting a Gaussian beam-shape to the data, giving a value for
CAL of 16.9$\pm$1.9\,K. Although this result is somewhat
higher than the value found using the Moon, the two results are
 consistent within the error.

\section{The maximum entropy reconstruction}
\label{MEM}

The stack shown in Fig.~\ref{Fig:MEMstack} has data points every 2 minutes in RA.
At Dec=+41\degr\ the interferometer beam pattern is $\sim$20 minutes wide
so there are $\sim$10 independent points per resolution element in each
of the
cosine and sine channels. By reducing the number of independent points
in each resolution element to 1, improvements to the signal-to-noise
ratio of the order of $\sim\sqrt{10}$ should be possible. Simple
averaging cannot be used to do this, as positive and negative lobes
would cancel. However by de-convolving the data with a technique such
as the maximum entropy method (MEM, Gull 1989) and then re-convolving the
result with the interferometer beam pattern, it is possible to obtain
one independent point per resolution element. The MEM algorithm
described in Maisinger \et (1997) was used to
simultaneously deconvolve cosine and sine data, using beam shapes of
$\exp[-{\rm RA}^2/2\sige^2] \cos (2\pi {\rm RA}/f)$ for the cosine
channel and $\exp[-{\rm RA}^2/2\sige^2] \sin (2\pi {\rm RA}/f)$ for the
sine channel. Here \sige and $f$ must be corrected to declination
+41\degr. The MEM-processed Dec=+41\degr\ stack can be seen in
Fig.~\ref{Fig:MEMstack}.

\subsection{Errors in the MEM reconstruction}

The MEM technique does not determine errors; we use Monte Carlo techniques to make an estimate
of the error in the MEM reconstruction using different signal-to-noise
ratios. These simulations were generated from a sum of cosines which populate
the spatial  frequencies to which the interferometer is sensitive, namely
\begin{equation} 
\phi=\sum_{i=5}^{300} \cos[2\pi\frac{\rm RA}{i/10} + \theta]
\end{equation}
where $\theta$ is a random phase between 0 and $2\pi$.
These were convolved with the interferometer beam pattern to produce
a noise-less sky which was normalized to 10\,$\mu$K RMS. White noise with
a known RMS was added to create data with signal-to-noise ratios between
1 and 5.

\begin{figure}
\leavevmode\epsfbox{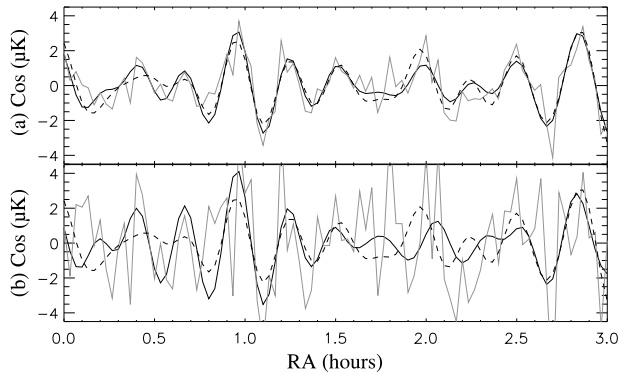}
\caption{\label{Montec} 
Example sky models used for Monte--Carlo MEM analysis. The noiseless
sky model, shown by dashed lines, has an RMS signal
level of 10~$\mu$K (comparable with expected CMB levels).
Different RMS levels of noise are added --
in these examples:
(a) 10\,$\mu$K and (b) 30\,$\mu$K.
The resulting simulated noisy data are shown with faint lines.
In each case a MEM analysis is performed. The resulting reconstructed sky
signals are shown with solid lines.}
\end{figure}

Two example simulated data sets are shown in Fig.~\ref{Montec}.
In the upper trace 10~$\mu$K (RMS) noise has been added
(a signal-to-noise ratio of 1:1).
These simulated data were de convolved using the same parameters as
those used for the real data.
The MEM reconstruction reproduces
the noiseless signal well. In the second case 30\,$\mu$K noise
has been added. This time the difference between the noiseless data
and the reconstruction is larger, but the process is still reproducing real
features from the data, despite the high noise level.

 The MEM process outlined in Maisinger \et (1997) also requires the
parameters $m$ and $\alpha$ which are respectively the default value
for the de convolved sky in the absence of any information, and a
regulating parameter which changes the relative weights given to
fitting the de convolved sky to the data, and maximising the entropy. As
indicated in Maisinger \et, varying $m$ and $\alpha$ by up to 2
orders of magnitude produced no noticeable change in the re-convolved
skies, thus demonstrating the reliability of MEM. 

\begin{figure}
\leavevmode\epsfbox{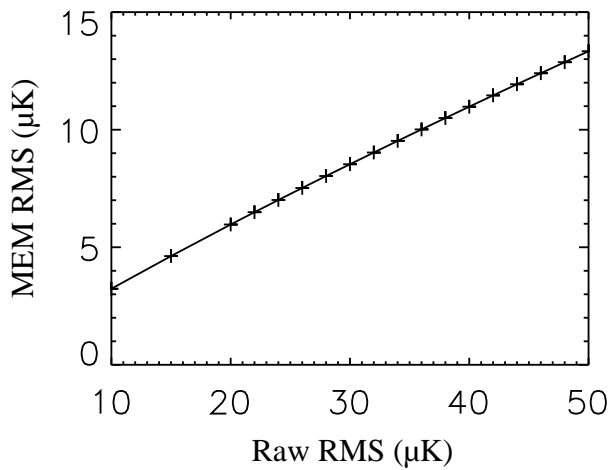}
\caption{\label{RMS} The RMS noise in the MEM-processed data over
the one beam width resolution element plotted against the RMS noise in the
2-minute data stack. Each point was found by de-convolving 100 simulated
skies.} \end{figure}

The RMS of the difference between each re-convolved
result and the corresponding noise-less observed signal
was calculated. This process was repeated over 100 skies. The average
RMS error in the re-convolved results plotted against the RMS noise
added to the simulated data can be seen in Fig.~\ref{RMS}. The best-fit
quadratic line to this graph was used to estimate the errors of the data
in Fig.~\ref{Fig:MEMstack} from the errors in the Dec=+41\degr\ stack. On
average MEM decreased the errors by a factor of 3.4 as expected from
the number of 2-minute points per resolution element.
By de-convolving the simulated skies containing 10\,$\mu$K of signal and
10\,$\mu$K noise it was shown that using  beam widths and fringe spacing up
to 5 per cent different to those used to create the fake data had little
effect on the re-convolved output from MEM.  As the errors in the
measured beam-shape are less than 2 per cent, they will have a
negligible effect on the output from MEM.

\subsection{Evidence for structure}

 The MEM-processed data show a good match to the unprocessed 
stack, especially over the Cygnus region.  The average RMS error on 
each point, as calculated using the above Monte Carlo simulations, is 
7.1\,$\mu$K and from 12\hr\ to 19\hr\ the RMS on some points is lower 
than 6.0\,$\mu$K. Many features in Fig.~\ref{Fig:MEMstack}
 are real at a greater than 2-sigma level and some 
(for example at 11\hr\ RA) are real at a 6-sigma level.  
The origin of these features is discussed in Section~\ref{dis} while 
their RMS amplitude can be estimated by the quadrature subtraction of
the uncertainty in the data from the 
RMS scatter of the data points (as estimated from Fig.~\ref{Fig:MEMstack}).
 Between 8\hr00 and 19\hr30 RA the cosine and sine data 
have RMS scatters of $12.2 \pm 1.4\,{\rm\mu K}$ and $12.1 \pm 1.4\,{\rm\mu K}$
 respectively.
The errors in these scatters take into account that neighbouring points 
are not independent. 
Over the same region the corresponding averages of the RMS records are 
$6.9 \pm 1.0\,{\rm\mu K}$ and $6.8 \pm 1.0\,{\rm\mu K}$ so the observed
astronomical signal  has a RMS amplitude of $10.1 \pm 1.8\,{\rm\mu K}$. Using 
the window function from Section~\ref{like} and assuming that the 
power spectrum of the sky fluctuations is flat, it can be shown that 
this corresponds to an intrinsic fluctuation amplitude of 
$\Delta T_\ell = 57 \pm 10\,{\rm\mu K}$.
A more quantitative 
measurement of this structure using Maximum Likelihood follows.

\section{Likelihood analysis}
\label{like}

Methods based on the likelihood function have been used extensively in
the analysis of CMB data (see for instance Hancock \et 1997). The
method considers the statistical probability distribution of the CMB
signal, and takes into account how the observing strategy, experimental
configuration, etc. modifies the statistical properties of the sky
signal. In standard models, the CMB fluctuations are described by a
Gaussian multi-normal random field, which is fully described by its
power spectrum or alternatively by the two-point correlation function.
In this case the likelihood function follows a multi-normal distribution
with the covariance matrix composed of two terms due to the signal and
noise respectively: $\bf C=S+N$. In our case the data consist of a set
of differences in temperature of the cosine and sine channels along
with their error bars binned in 2-minute bins in RA. For the noise
correlation matrix, the only non-zero terms are those on the diagonal,
as the instrumental noise is uncorrelated from point to point.

\begin{figure}
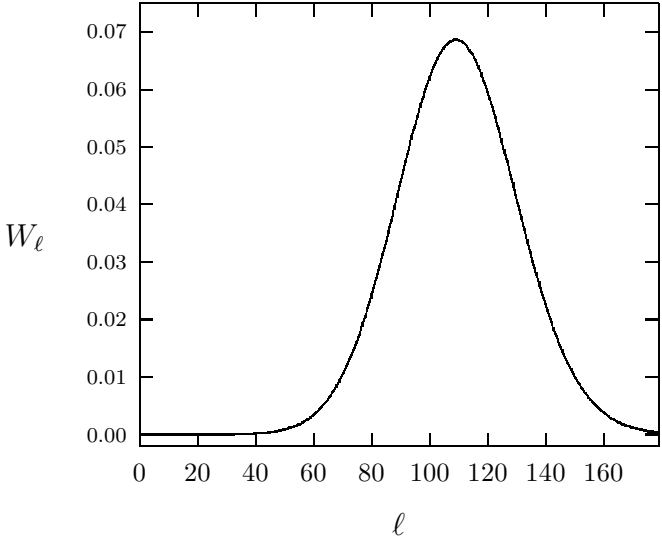
  
\include{fig9}
\caption{\label{window}The window function of the interferometer.}
\end{figure}

We have assumed that the CMB signal has a flat power spectrum ($\Delta
T_\ell\equiv \sqrt{l(l+1)C_\ell/2\pi}$ constant) over the range covered by
the window function of the instrument, and then the expected two-point
correlation between a pair of points $i$ and $j$ separated by an angle
$\theta_{ij}$ is given by 
\begin{eqnarray}
\left<T_i\cdot T_j\right>\  & = &
\exp\left(\frac{-\theta_{ij}^2}{4\sigma^2}\right)
\cos\left(\frac{2\pi\theta_{ij}} {f}\right)\frac{\Delta T^2_{l}}{2}
\nonumber \\
& \times & \sum _{l_1}^{l_2} \frac{2l+1}{l(l+1)}W_\ell(\theta _{ij})
\end{eqnarray}
$W_\ell(\theta_{ij})$ being the window function of the experiment which defines
the response of the instrument to a given multipoles or angular scale.
The computation of the window function of an interferometer like the
one analyzed here is not straight-forward. We adapted the method
of Muciaccia, Natoli and Vittorio (1997) to
decompose the beam configuration into spherical harmonics,
which can then be used to form the window function.
The resulting function (see Fig.~\ref{window}) can be fitted by:
\begin{equation}
W_{\ell}(0) =
0.677\exp(-(\ell-109)(\ell-110)/730).
\end{equation}
This defines the range of
sensitivity to be $l\sim 109\pm19$. 
We use this to calculate the expected excess variance
in the data for some theoretical  model prediction on band power \cite{Cite:Bond97}.
Starting with the convolved sky covariance function at zero lag, which is the
same as the data variance
\begin{equation}
C_m(0)\equiv \frac{1}{4\pi}\sum_\ell{(2\ell+1)C_{\ell}W_{\ell}(0)},
\end{equation}
we can substitute for $C_\ell$ from the definition of band power to obtain
the ratio of filtered to non-filtered variance
\begin{equation}
C_m(0)/\Delta T_{\ell}^2 = \frac{1}{2} \sum_\ell
\frac{(2\ell+1)}{\ell(\ell+1)}W_\ell(0) = 32.5
\end{equation}

For the likelihood analysis it is necessary
to construct the expected theoretical covariance matrix which
corresponds to carrying out the above procedure for non-zero lag
window functions. As the window is narrow we can just use the
beam autocorrelation function and normalize it to the ratio found above for
zero lag.

\subsection{The Galactic cut}

To locate regions free of significant
Galactic emission, 48 intervals of 5 hours in RA each stepped
by 30 minutes in RA were analysed using
the likelihood function. It was found that the amplitude
of the detected signal was at a low stable level in the range RA=8.0--19.5
hours, indicating a low level of foreground contamination over this
region showing no change with Galactic latitude. 
Further indications that this
region is free from foreground
contamination are discussed in Section~\ref{Gal}. It is worth noting
that when data from the sine channel for the region RA=8-11 hours are
included
in the analysis there is marginal evidence (at the one-sigma level) of a
slightly larger signal as compared with the results in RA=11.0--19.5
hours and those of the cosine channel; this effect could be due to some minor
atmospheric residual in this channel (see Section \ref{basic} and below).

\subsection{The results}

A likelihood analysis of each channel over the RA=8.0--19.5 hour
range gave $\Delta T_\ell=49^{+12}_{-11}$ and
$37^{+18}_{-18}$ $\mu$K (68 \% C.L.) for the cosine and sine channels
respectively. The larger uncertainty for the sine channel reflects the shape
of the likelihood function which has lower peak and is broader than the
likelihood function of the cosine channel, however the signals detected in
each channel are consistent at the one-sigma level. An additional test
of the consistency and repeatability of these results was carried out
by the likelihood analysis of two independent data stacks, one formed
from the first 50 days of data used to make the main stack and another
formed from the remaining data. The results for each sub-stack are in
agreement with each other and with  the above results obtained when the
full data of each channel are analyzed. 

The best estimation of the signal present in our data comes from the
joint analysis of both channels. When data from both the cosine and sine channels are
analyzed together building the joint likelihood function the signal
detected is $\Delta T_\ell=45^{+13}_{-12}$ $\mu$K at the 68 \% C.L.
This indicates not only that the amplitude of the signal detected
in each channel is in agreement, but also that the signal detected by
both channels comes from the same structures.

With a limited sky coverage consideration of the
``sample variance'' \cite{Cite:ScottSredWhite94} is
crucial to an understanding of the quoted errors.
To quantify this contribution we integrate the square of the telescope
two-point correlation function over the survey area
(equation 4 of Scott et al.).
For our single 11.5-hour RA strip ($\sim$35 independent beam widths)
the contribution is 20.7 per cent, or $8.9\,{\rm\mu K}$.
In comparison the receiver noise contribution is approximately $5.2\,{\rm\mu K}$,
added in quadrature (see Table \ref{Table:contrib}).
The maximum likelihood error is slightly larger than expected, possibly
due to a contribution from weather effects.
The effect of finite sky coverage represents the majority of the error in our determination
of the amplitude of CMB fluctuations. 
Additional observations at other declinations, which are being conducted now, 
will increase our sky coverage, significantly reducing this uncertainty.

The contributions of weak point sources and Galactic emission, which would add in 
quadrature to the $\Delta T_\ell$ arising from the CMB, are discussed in sections 
\ref{point} and \ref{Gal}, below.
Since these could lead to a small over-estimate in $\Delta T_\ell$, we
extend the negative error bar slightly, as given in Table \ref{Table:contrib}.

\begin{center}
\begin{table}
\caption{Contributions to the error in $\Delta T_\ell$.
The receiver noise and sampling error add in quadrature,
accounting for most of the maximum liklihood error.
Weak point sources and Galactic emission would contribute in quadrature to $\Delta T_\ell$
and so can only have lead to a small over-estimate.
This is allowed-for by a small extension to the negative error on $\Delta T_\ell$.
The survey is referenced to the Moon temperature,
with the calibration uncertainty given.}
\label{Table:contrib}

\begin{tabular}{llrlr}
 Receiver noise  & &  $\pm$ & $ 5.2$ & ${\rm\mu K}$  \\
Sampling error & (20.7\,\%)   & $\pm$ & $8.9$ & ${\rm\mu K}$  \\
Weak point sources & ($\Delta T \la 8\,{\rm\mu K}$) & $-$ & $0.5$ & ${\rm\mu K}$  \\
Galactic free-free & ($\Delta T \la 2\,{\rm\mu K}$) &  $-$ & $0.05$ & ${\rm\mu K}$  \\
\hline
\multicolumn{2}{l}{Maximum Likelihood error}   &    &  $^{+12.5}_{-12.0}$ & ${\rm\mu K}$\\
\hline
{\footnotesize Calibration}  &   &  {\footnotesize $\pm$} & {\footnotesize $ 6.6$} & {\footnotesize \%}  \\
\end{tabular}
\end{table}
\end{center}

\section{Foregrounds}
\label{dis}

\subsection{Point Sources}
\label{point}

\begin{center}
\begin{figure*}
\leavevmode\epsfbox{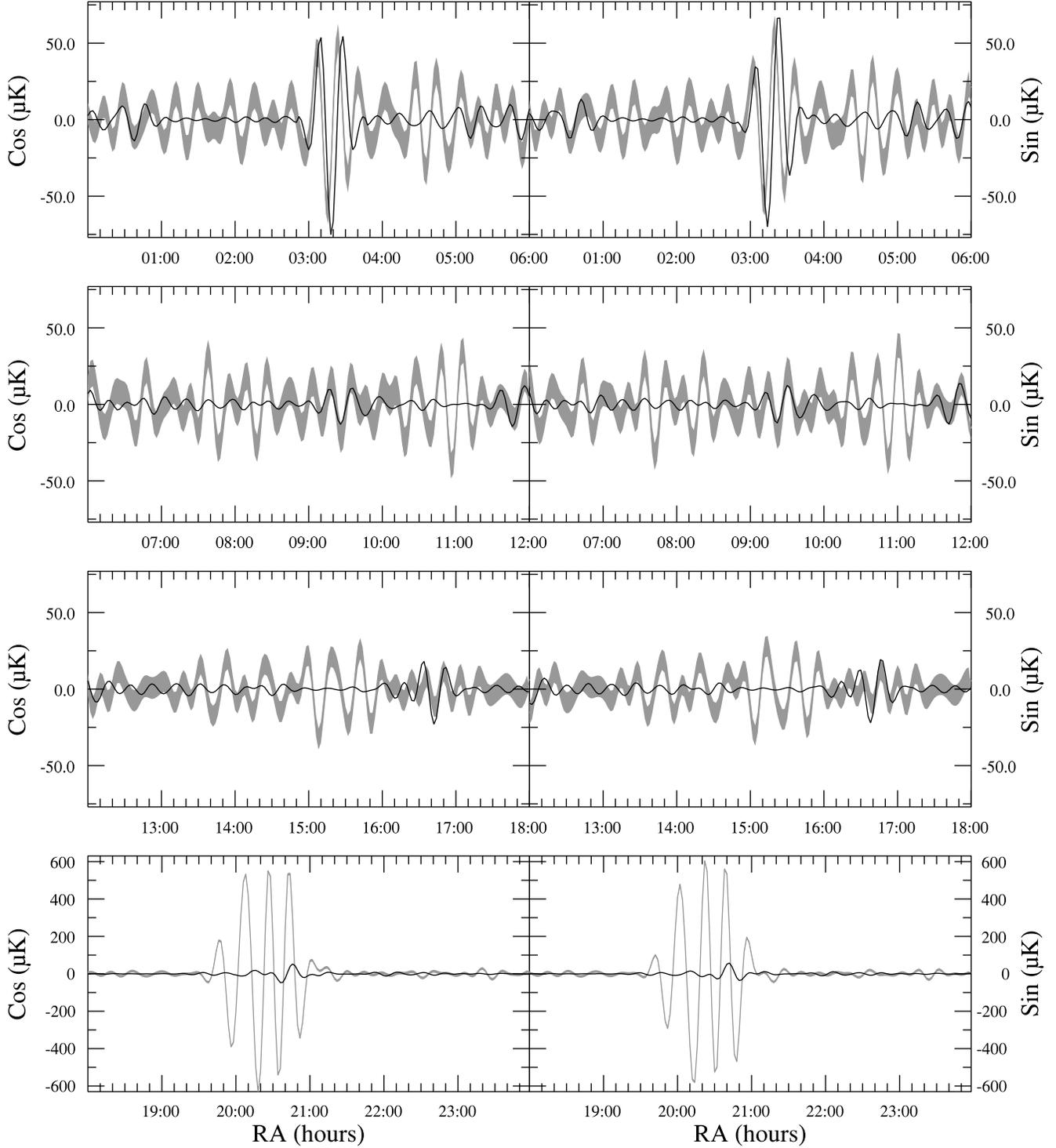}
\caption{\label{conv}The contribution of point sources with predicted 
S(33GHz)$\ge$0.2\,Jy in the field of the Dec=+41\degr\ scan. The black lines
represent the predicted point source contribution, while the grey lines are the
MEM-fitted data with their 1$\sigma$ errors.}
\end{figure*}
\end{center}

Only the strongest sources at 33\,GHz have reliably measured flux densities
since no large-area survey is available at this frequency. Furthermore, many
of the strongest sources have flat spectra and are time-variable. The 5
sources with S(33\,GHz)$\ge$ 2\,Jy within the 4$^\circ$-wide strip centred on
Dec=41\degr\ are monitored on a continuous basis in the Metsahovi 22 and
37\,GHz programme (Ter\"asranta \et 1992). They are listed in
Table~\ref{sources}. The sources are all variable; Dr. Harri Ter\"asanta has
kindly provided data covering the observing period of the present CMB survey.

\begin{table}
\caption{Sources within a 4$^\circ$-wide Dec strip centred on +41\degr.}
\label{sources}
\begin{tabular}{@{}l|cc}
       Name &   RA$_{\tiny 2000}$        &  Dec$_{\tiny 2000}$\\
\hline
     3C 84  & 03\hr\  19\mn\  48\rs & +41\degr\ 30\arcmin\ 42\arcsec \\
    DA 193  & 05\hr\  55\mn\  31\rs & +39\degr\ 48\arcmin\ 49\arcsec \\
   4C 39.25 & 09\hr\  27\mn\  03\rs & +39\degr\ 02\arcmin\ 21\arcsec \\
   3C 345   & 16\hr\  42\mn\  59\rs & +39\degr\ 48\arcmin\ 37\arcsec \\
   BL Lac   & 22\hr\  02\mn\  43\rs & +42\degr\ 16\arcmin\ 40\arcsec \\
\hline
\end{tabular}
\end{table}

A limited amount of 33\,GHz data on other (weaker) sources is available in the
Kuhr \et (1981) Catalogue. The highest radio frequency survey covering the sky
around Dec=+41\degr\ is the 4.85\,GHz Green Bank survey (Gregory \et 1996).
Those sources with S(5\,GHz)$\ge$0.2\,Jy lying within Dec=41\degr$\pm$3\degr\
were selected as possible contributors to the 33\,GHz point source background.
The Kuhr catalogue gives measured flux densities in the range 14--37\,GHz for
90 per cent of the Green Bank sources with S(5\,GHz)$\ge$1\,Jy. For these, a
reliable 33\,GHz flux density could be derived. For the remaining sources with
S(5\,GHz)$\ge$0.2\,Jy, S(33\,GHz) was estimated using the spectral index between
the flux density given in the 1.4\,GHz NVSS Catalogue (Condon \et 1989) and
that given in the Green Bank 5\,GHz survey, assuming a spectrum of the form
$S\sim\nu^\alpha$ where $\alpha$ is the spectral index. Where no matching
1.4\,GHz source was found, a conservative index of -0.1 was assumed.

The 33\,GHz flux densities of the sources identified above were then convolved
with the two-dimensional interferometer beam pattern centred on Dec=+41\degr.
The flux densities were converted to antenna temperature using the factor
7.07$\mu$K\,Jy$^{-1}$ calculated in Section~\ref{tau2}. The predicted contribution of
these point sources is compared with the MEM-reduced observed data in
Fig.~\ref{conv}. The sources 3C84 and 3C345 show good matches to the
observed data. Over the Galactic plane regions (RA=5\hr30\mn\ -- 6\hr30\mn\
and 19\hr30\mn\ -- 21\hr30\mn) flux predictions for point sources were
difficult as most surveys avoid these crowded areas. In these regions
we expect most of the observed structure to be due to Galactic
sources and, as we have excluded these regions from our analysis to
determine the CMB structure, the contribution from both Galactic and
extra-galactic sources is irrelevant.

Excluding the Galactic regions and the 5 strongest point sources, the RMS of
the MEM-reduced data is approximately 10 times that predicted from
discrete point sources alone. As a check that  the likelihood results do not
depend strongly on the point source prediction, the above likelihood
analysis was repeated with and without the subtraction of point
sources. It was found that with the point sources subtracted the result
of the analysis of the cosine and sine data together gave $\Delta
T_\ell=43^{+12.5}_{-11.5}\,{\rm\mu K}$ as compared to $\Delta T_\ell=45\,{\rm\mu K}$ without,
thus demonstrating that point source contribution is not a major concern for
the data presented here.

Although it has been shown that the discrete resolved point sources do not
contribute significantly to our observed value of $\Delta T_\ell$, it is 
necessary to quantify the contribution due to a foreground of unresolved 
point sources.
The expected RMS for a random distribution of such sources varies with
frequency and angular scale.
On scales of 1\fdg6 and at a frequency of 33\,GHz, Franceschini \et
(1989) predict that unresolved point sources should  have \TT\ $\approx
3\times10^{-6}$ corresponding to $\Delta T = 8\,{\rm\mu K}$. 
This would add a contribution in quadrature, accounting for approximately 
$0.5\,{\rm\mu K}$ of the total.
Unresolved point sources are therefore unlikely to make a significant
contribution to our estimate of $\Delta T_\ell$.
We may allow for this component by extending the negative error range
by $0.5\,{\rm\mu K}$, giving $\Delta T_\ell=43^{+12.5}_{-12.0}\,{\rm\mu K}$, 
as given in Table \ref{Table:contrib}.

\subsection{Diffuse Galactic emission}
\label{Gal}

An indirect estimate of the amplitude of the diffuse Galactic component
in our data can be computed using the results obtained in the same region of the sky by
the Tenerife CMB experiments (Guti\'errez \et 1998). At 10.4\,GHz and on
angular scales centred on $\ell = 20$ the maximum Galactic component was
estimated to be $\le 28\,{\rm\mu K}$. Assuming that this contribution is
entirely due to free-free emission and a conservative Galactic spatial
power spectrum of $\ell^{-2.5}$ (Lasenby 1996, Davies \& Wilkinson 1998), the
predicted maximum
Galactic contamination in the data presented here is $\le 2\,{\rm\mu K}$, less
than 5 per cent of our measured value. 
Any such contribution would add in quadrature to that from the CMB,
accounting for approximately $0.05\,{\rm\mu K}$, 
which is insignificant.
The true make-up of the Galactic 
foreground emission detected at 10.4\,GHz most probably has a steeper 
average index than this, so the
contribution to our result will be even lower than stated. 

Several experiments \cite{Cite:Kogut,Cite:Oliveira97,Cite:Leitch}
have detected an ``anomalous'' microwave emission,
correlated with 100-${\rm\mu m}$ thermal emission from interstellar dust.
It has been argued \cite{Cite:DraineLaz98} that spinning dust grains
containing $\sim 10^2 - 10^3$ atoms might be responsible. Our survey
region is in an area of low 100-${\rm\mu m}$ emission, but even so 
a contribution from spinning dust comparable to that of free-free emission
is possible. Again, such a contribution would add in quadrature to the CMB,
and even a level as high as $15\,{\rm\mu K}$ would account for
only $3\,{\rm\mu K}$ of our measurement.

\section{Conclusions and the future programme}
\label{con}

We describe in this paper the results from the first high sensitivity data
stack taken with the new Jodrell~Bank~--~IAC 33\,GHz interferometer in a strip
at Dec=+41\degr. By using the MEM technique it has been possible to reduce the
RMS noise in each 5\degr\ RA beam width of the stack to $6\,{\rm\mu K}$. The
source 3C84 is clearly seen in the data at the level expected from its
monitored intensity at Metsahovi. All the other individual sources are weaker
than the measured signals in the raw data as shown in Fig.~\ref{conv}.
Furthermore,  arguments
are presented in Section~\ref{Gal} that the RMS Galactic contribution,
neglecting dust, is
$\Delta T \le 2\,{\rm\mu K}$, and is therefore negligible. Thus, apart from the
region of the stack around 3C84 (RA=3\hr 20\mn) and the strong Galactic plane
crossing (RA=19\hr.5--21\hr.5) the stack shows significant CMB signals over the
entire RA range. Many of these CMB features have amplitudes 4--5 times the RMS
noise in the 5\degr\ beam width. When corrected for dilution in the beam of the
interferometer these features have a sky brightness temperature of 
$\sim 100\,{\rm\mu K}$.
\begin{figure}
\leavevmode\epsfbox{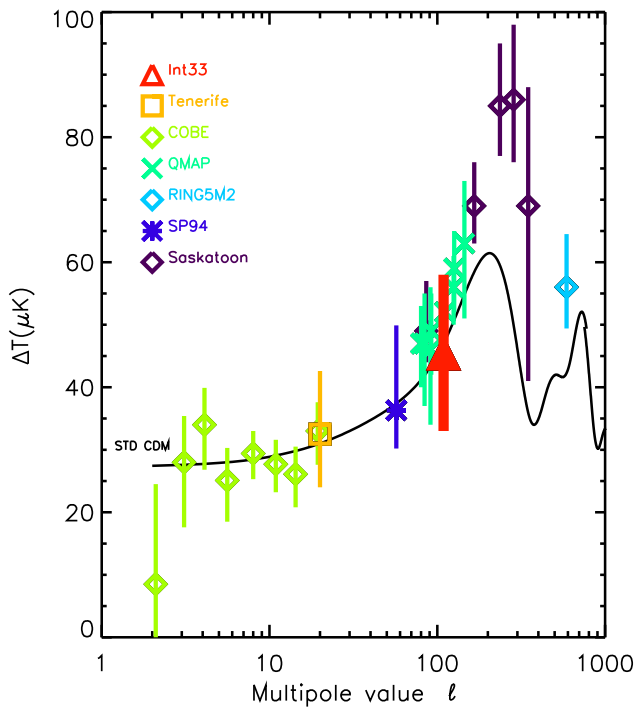}
\caption{\label{Clplot}The C$_\ell$ versus $\ell$ plot of recently published data
on CMB anisotropies including the present result (shown by the heavy line) of
$\Delta T_\ell=43^{+12}_{-12}{\rm\,\mu K}$ at $\ell = 109 \pm 19$.}
\end{figure}

The value of the intrinsic CMB fluctuation amplitude derived from the present
observations at high Galactic latitudes is 
$\Delta T_\ell = 43^{+12.5}_{-12.0}\,{\rm\mu K}$.
In the
maximum likelihood analysis only one 11.5-hour long RA strip is used. This   
leads to a sampling error of approximately 21 per cent,
which is the main contributor to the $\sim 25$ per cent uncertainty of our result.
The remainder is due to receiver noise. 
Clearly the major improvement in an
estimate of $\Delta T_\ell$ will come from the coverage of a larger area of the
sky
at the present or better receiver sensitivity. Further observations are being
undertaken at adjacent declinations of +39\fdg8 and 42\fdg2 in addition to the
full RA range at Dec=+41\degr. These should increase the area covered by a factor
of 4 which will reduce the sampling error to approximately 10 per cent,
comparable with the receiver noise contribution, reducing the total error to $\sim$ 15 per cent.
In addition to the quoted error there is a calibration uncertainty of $\pm 6.6$ per cent,
arising mostly from uncertainty in the Moon temperature. Future improvements
to the calibration are possible.

This result obtained at 32.5\,GHz and at high Galactic latitude may be compared
with others at similar values of $\ell$ but made at different frequencies and
Galactic environments. Our best estimate is 
$\Delta T_\ell = 43^{+12.5}_{-11.5}\,{\rm\mu K}$
in the range $\ell = 109\pm19$. Two such recent measurements have been made in
the North Celestial Pole (NCP) region.
One is the Saskatoon experiment (Netterfield \et 1997) at
26--46\,GHz which found $\Delta T_\ell=49^{+8}_{-5} {\rm~\mu K}$ at $\ell = 87$; there
is a further 15 per cent calibration uncertainty in this result. The other is
QMAP (De Oliveira-Costa \et 1998) which at 30\,GHz found
$\Delta T_\ell=47^{+6}_{-7}\,{\rm\mu K}$ at $\ell = 80$ and
$\Delta T_\ell=59^{+6}_{-7}\,{\rm\mu K}$ at $\ell = 126$ while at 40\,GHz it found 
$52^{+5}_{-5}\,{\rm\mu K}$ at $\ell = 111$. At the South Celestial Pole (SCP),
the SP94 (Gundersen \et 1995) estimated $\Delta T_\ell=36^{+13}_{-6}\,{\rm\mu K}$
at $\ell = 60$ and Python (Platt \et 1997) obtained  $\Delta
T_\ell=60^{+15}_{-13}\,{\rm\mu K}$ at $\ell = 87$. These values, along with the result
reported here and other recently published values, are plotted on
Fig.~\ref{Clplot}. All these results together strongly indicate a significant
increase in fluctuation amplitude at $\ell\sim 100$ compared with the COBE value
of $\Delta T_\ell=30\,{\rm\mu K}$ at $\ell\sim 10$, thereby arguing for the
existence of the first Doppler peak.

Another future contribution by the interferometer will be observations on
a scale of $\sim$ 1\degr\ ($\ell \sim$ 200) with a baseline of 304\,mm in place of
the present 152\,mm. Such observations will enable us to sample the first
acoustic peak. The values of $\Delta T_\ell$ at $\ell = 100$ and 200 will thus be
directly compared using the same instrument and the same calibration methods,
enabling a better estimate to be made of the width and amplitude of the first
peak.

\section{Acknowledgements}

This work has been supported by the European Community Science program
contract SCI-ST920830, the Human Capital and Mobility contract
CHRXCT920079 and the UK Particle Physics and Astronomy Research
Council. AW acknowledges the receipt of a Daphne Jackson Research Fellowship,
SRD and DLH PPARC Postgraduate Studentships.
We thank Dr.H.Ter\"asranta for providing data on point sources at 22 and 37\,GHz.

\label{lastpage}

\end{document}